\documentclass[epjCONF]{svjour}
\usepackage[pdftex]{graphics}
\usepackage[varg]{txfonts} 
\usepackage[latin1]{inputenc}
\pdfoutput=1
\usepackage{natbib}
\session-title{A Universe of Dwarf Galaxies}
\begin{document}
\title{UCDs in the Coma Cluster}
\author{Kristin Chiboucas\inst{1}\fnmsep\thanks{\email{kchibouc@gemini.edu}} \and R. Brent Tully\inst{2} \and Ronald O. Marzke\inst{3} \and Steven Phillipps\inst{4} \and James Price\inst{4} \and Eric W. Peng\inst{5} \and Neil Trentham \inst{6} \and David Carter\inst{7} \and Derek Hammer\inst{8}}
\institute{Gemini Observatory, 670 N. A'ohoku Pl, Hilo, HI 96720, USA \and Institute for Astronomy, University of Hawaii, 2680 Woodlawn Dr., Honolulu, HI 96821, USA \and Department of Physics and Astronomy, San Francisco State University, San Francisco, CA 94132, USA \and Astrophysics Group, H.H. Wills Physics Laboratory, University of Bristol, Tyndall Avenue, Bristol BS8 1TL, UK \and   Kavli Institute for Astronomy and Astrophysics, Peking University, Beijing 100871, China  \and Institute of Astronomy, Madingley Road, Cambridge CB3 0HA, UK \and Astrophysics Research Institute, Liverpool John Moores University, Twelve Quays House, Egerton Wharf, Birkenhead CH411LD, UK \and Department of Physics and Astronomy, Johns Hopkins University, 3400 North Charles Street, Baltimore, MD 21218}
\abstract{
As part of the HST/ACS Coma Cluster Treasury Survey, we have undertaken a Keck/LRIS 
spectroscopic campaign to determine membership for faint dwarf galaxies. In the process, 
we discovered a population of Ultra Compact Dwarf galaxies (UCDs) in the core region of 
the Coma cluster. At the distance of Coma, UCDs are expected to have angular sizes 
$0.01 <$ R$_e < 0.2$ arcsec. With ACS imaging, we can resolve all but the smallest ones 
with careful fitting. Candidate UCDs were chosen based on magnitude, color, and 
degree of resolution. We spectroscopically confirm 27 objects as bona fide UCD 
members of the Coma cluster, a 60\% success rate for objects targeted with $M_R < -12$. 
We attribute the high success rate in part to the high resolution of HST data and to 
an apparent large population of UCDs in Coma. We find that the UCDs tend to be strongly 
clustered around giant galaxies, at least in the core region of the cluster, and have a 
distribution and colors that are similar to globular clusters.  These findings suggest
that UCDs are not independent galaxies, but rather have a star cluster origin.
This current study 
provides the dense environment datapoint necessary for understanding the UCD population. 
} 

\maketitle

\section{Introduction}
\label{intro}
Ultra compact dwarf galaxies (UCDs) were first discovered only a decade ago in the nearby
Fornax cluster \citep{hil99, djgp00, phill01}.  Their compact sizes and broad range of colors make 
them nearly indistinguishable from stars in ground-based data.  To date, UCDs and candidate UCDs have 
been discovered in several other nearby clusters along with a handful that have turned up in very 
poor environments \citep[see e.g.][]{jones06,mieske07,wehner07,mieske06,hau09, norris10}.
None, up to this point, had been confirmed in as rich and evolved an environment as the Coma cluster.  

Post-discovery, 
attention has turned to revealing the nature of these unusual objects which have sizes in between
those of globular star clusters and dwarf galaxies, unusually red colors, high M/L ratios which
cannot be explained with canonical IMFs + baryonic matter, and, within the framework of certain
models, difficult to explain ages and metallicities.  Hypotheses in contention include formation
via threshing whereby a nucleated dwarf elliptical is disrupted in multiple orbits by the 
strong tidal forces of a cluster or giant galaxy, such low surface brightness dE,N that the envelope
is nearly invisible, formation of a giant star cluster from the mergers of super star clusters, and the simple
extension of the bright tail of the globular cluster luminosity function to exceptionally bright magnitudes 
\citep{djgp00,fk02,bekki03,has05,chil08}.

We have undertaken a Keck/LRIS spectroscopic survey of dwarf galaxies in the Coma cluster as part of the HST/ACS
Coma Cluster Treasury Survey \citep{carter}.  As part of this survey, we have conducted a search for UCDs in this
very dense, rich, evolved environment.  The aim has been to establish whether a population exists, and if so,
characterize the properities, population size, and distribution in order to better understand the
origin of these enigmatic objects.  

If UCDs are formed from the threshing of dE,N, one would expect
to find them in greater numbers than in lower density environments and with broader distributions, 
yet remain concentrated towards the cluster core.  If instead, they are simply glorified globular clusters,
one would expect to find physical properties and a distribution similar to that of the Coma cluster
globular cluster population.  They should also be associated with individual galaxies.  A super star 
cluster merger scenario would predict that UCDs be associated with galaxies that had undergone major 
mergers and that they share similar metallicities and ages with other stars in the galaxy that 
formed during the same event. A star cluster origin would also predict 
enhanced alpha element ratios due to the expected short timescale of star bursts that produce star clusters.
For an environment like the Coma cluster, the formation from the mergers of super star clusters, themselves 
formed during major galaxy mergers, would be unlikely.  Instead these would have to be old populations, 
created in the smaller galaxy groups that eventually formed the Coma cluster.

\section{The UCDs}
\label{sec:1}
\begin{figure}
\resizebox{1.05\columnwidth}{!}{\includegraphics{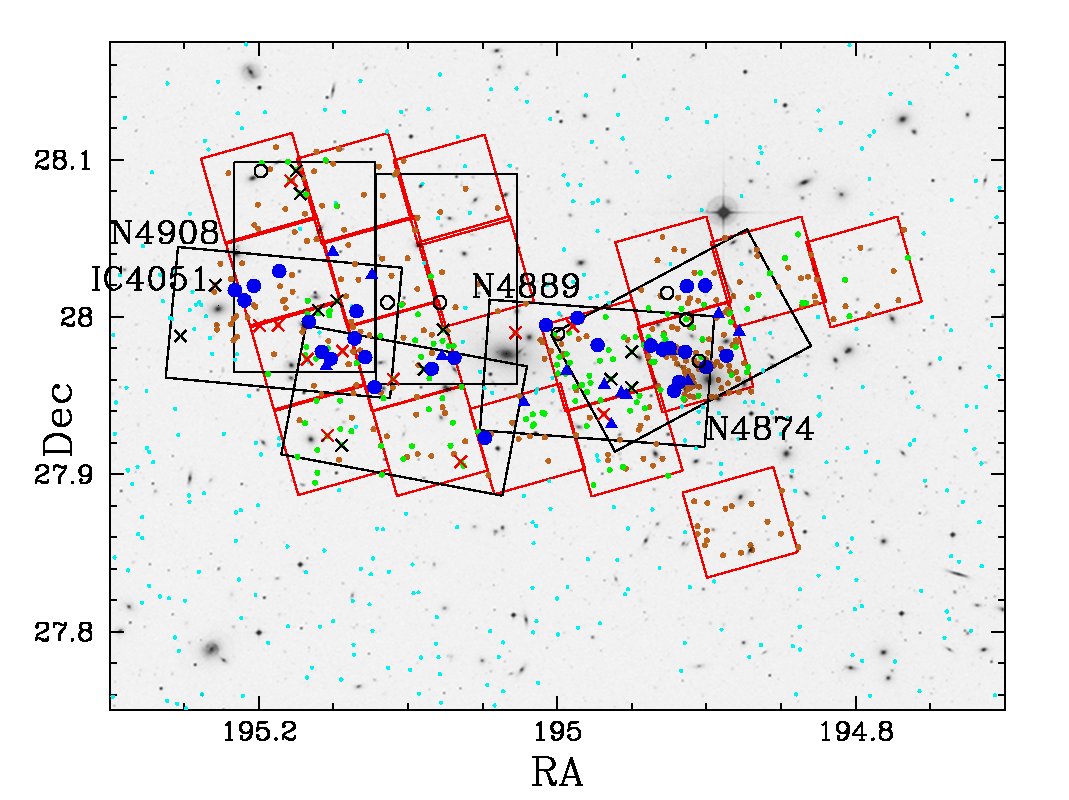}}
\caption{Spatial distribution of the UCDs. Red boxes are the locations of observed 
ACS fields, black boxes are the 6 LRIS masks. Green points represent our original candidate sample,
brown points are an expanded candidate sample, and cyan points are candidates chosen
strictly on the basis of color.
Larger blue circles denote the location of confirmed UCDs, triangles are 
uncertain UCDs from low S/N spectra. Open circles mark the location of compact 
dEs \citep{compact}. X's are
UCD candidates determined from redshifts to be stars (black) and
background galaxies (red).
\label{fig1}}       
\end{figure}

Candidate UCDs were initially chosen based on fairly broad color criteria ($0.45 < (B-V) < 1.1$
and $0.15 < (R-I) < 0.6$). 
Since UCDs have typical sizes  $7 <$ r$_e < 100$ pc corresponding to $0.01 - 0.2$ arcsec at the distance
of the Coma cluster, the larger ones are just resolved.  For this original sample we gave higher weight to those
sources showing any sign of having a profile broader than a pure PSF.  In an initial observing run,
we targeted 47 candidates in 4 masks and confirmed 19 UCDs.  Ten objects proved to be
background galaxies or stars, while the remainder had too low S/N to measure reliable redshifts.  
For targets brighter than our completeness limit ($R < 23.3$), we find 66\% of this sample are cluster members!
A subsequent observing run with looser color and resolution criteria, and plagued by poor weather,
turned up an additional 8 UCDs.  This high success rate likely indicates the existence of a large population
of UCDs in the Coma cluster, at least within the core region, although part of this success rate can also
be attributed to the high spatial resolution of the ACS data.
The locations of the 27 confirmed UCDs along with candidate UCDs are displayed in Figure ~\ref{fig1}.  Objects
which proved to be stars or background galaxies are also shown. 

\subsection{Properties}
\label{sec:1a}
\begin{figure}
\resizebox{1.05\columnwidth}{!}{\includegraphics{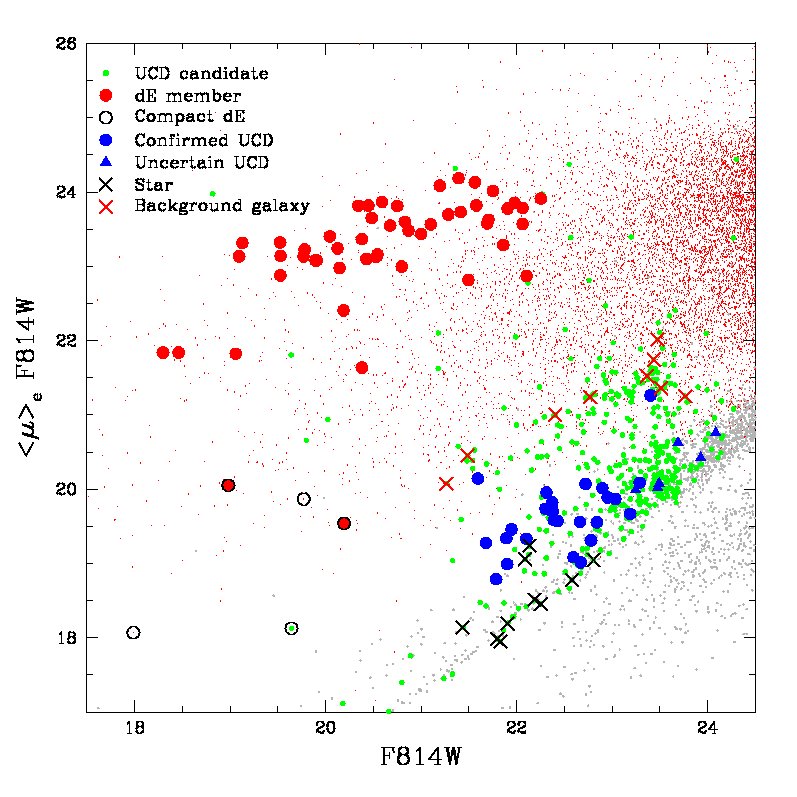}}
\caption{Mean effective surface brightness vs. $F814W-$band total magnitude
for all objects in our ACS survey region.  Photometry comes from
\citet{hammer}.
\label{fig2}}       
\end{figure}

A magnitude - surface brightness diagram (Figure ~\ref{fig2}) showing both confirmed dE/dE,N cluster members 
and confirmed UCDs turns up two nearly parallel sequences, at the extremes in surface brightness.  
The dEs follow the well known dwarf galaxy magnitude - surface brightness relation while the UCDs fall near 
the stellar locus.  However, unlike what comes out of ground-based data, the UCDs are separated from the 
stellar sequence, albeit with some
overlap.  Between these two cluster populations is largely the realm of background galaxies, although a few other
members are noted.  These are the compact dwarf ellipticals (cE) \citep{compact}, which either form a third
population of member galaxies, or follow a continuous sequence with the UCDs.  The latter possibility 
hinges on filling the 1.5 magnitude gap between the two populations, which we have thus far
been unsuccessful in accomplishing. 

\begin{figure}
\resizebox{0.8\columnwidth}{!}{\includegraphics{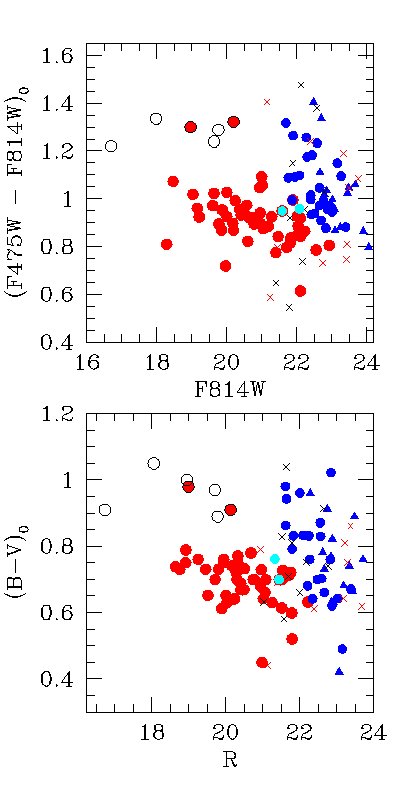}}
\caption{Colors for objects in our spectroscopic sample.
$B - V$ colors come from \citet{adami}, $F475W-F814W$ from our ACS
imaging.  Extinction is derived from the dust maps of \citet{sfd98}.
Confirmed UCDs and cEs \citep{compact} are large blue and open circles
respectively.  Triangles are possible UCD members with uncertain redshift measurements
due to low S/N spectra.
Cyan points are near-stellar
objects exhibiting possible faint low surface brightness envelopes.
Photometry does
not exist for all objects in all bands.
\label{cmd}}       
\end{figure}

Color-magnitude diagrams (Figure ~\ref{cmd}) find a remarkably wide spread in UCD colors.  
The red sequence of normal
dwarf galaxies is apparent, and a parallel but redder sequence of cE galaxies is noted.  The UCDs
meanwhile exhibit a spread of colors ranging between these two sequences and having a total extent of
$\sim0.4$ magnitudes in $F475W - F814W$ or $B-V$.

%

\begin{figure}
\resizebox{1.05\columnwidth}{!}{\includegraphics{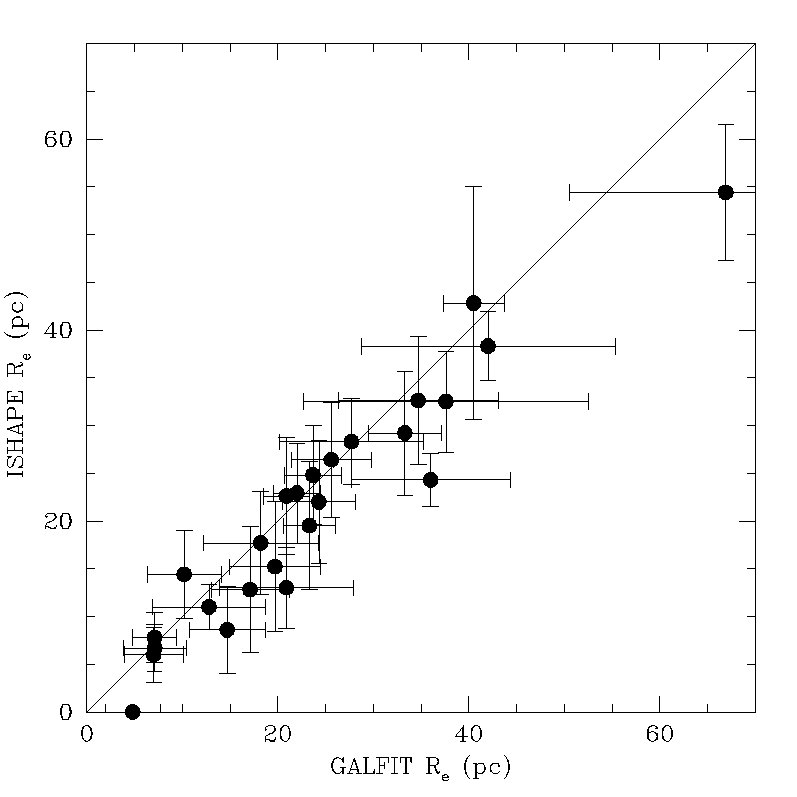}}
\caption{Comparison of Galfit and ishape measured R$_{e}$. 
One object has a much larger size and is not shown here.
\label{sizes}}       
\end{figure}

We have measured the sizes of the UCDs using the profile fitting software Galfit
\citep{peng02} and ishape \citep{larsen}.
The former is used to fit Sersic profiles while the latter, a program ideal for fitting marginally resolved
globular clusters, is used to fit King profiles.   To convert angular sizes to physical effective radii, we assume a distance
to the Coma cluster of 100 Mpc.  The two measurements agree very well (Figure ~\ref{sizes}) and sizes are found
to range between $\sim 7 - 40$ pc.  Two objects are found with significantly larger sizes, but residuals
from the fits show possible evidence for faint extended envelopes surrounding the objects.

\begin{figure}
\resizebox{1.05\columnwidth}{!}{\includegraphics{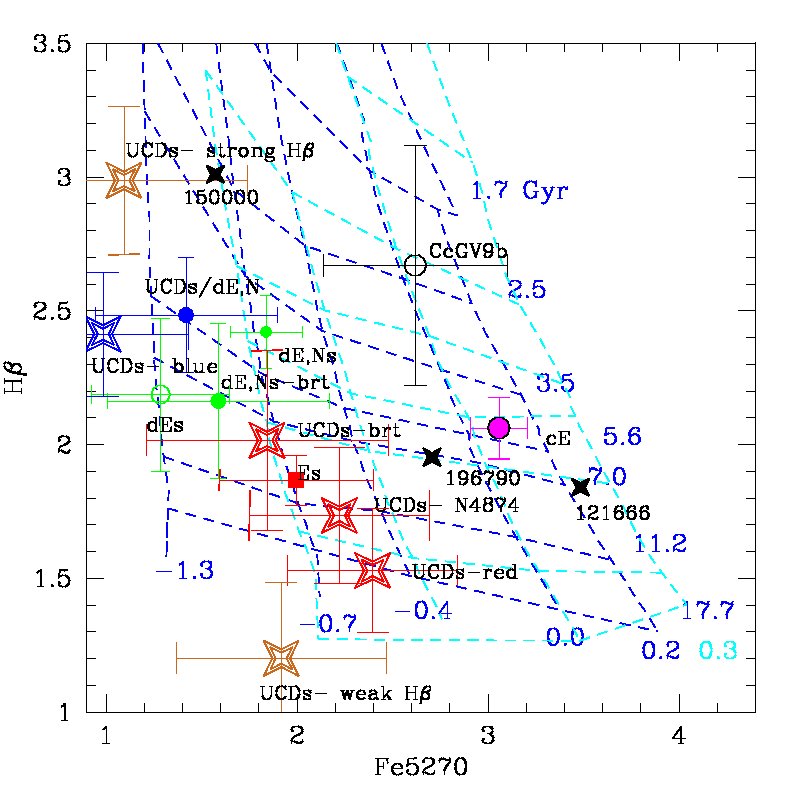}}
\caption{Measured Fe $5270\AA$ vs H$\beta$ line strengths.
[Fe/H]-age grid models are from \citet{schiavon}
for $\alpha/Fe = 0$ (blue) and 0.3 (cyan).
Plotted are Fe 5270 and H$\beta$ measurements for composite spectra
of similar object types.  Object types are labeled on the plot.  We include two points for
dE,N: the larger solid green circle includes 4 objects with bright, prominent nuclei, the smaller
one consists of 24 dE,N with small, faint nuclei.  The open black circle comes from \citet{compact} for
the same object labeled as cE below it.   Star-like symbols refer to UCDs with different
combinations of stacked spectra:
the 5 brightest UCDs, 6 red ($V-I > 1.05$),
9 blue ($V-I < 1.05$), 5 red UCDs around NGC 4874, 8 with weak H$\beta (< 2.5)$, and
8 with strong H$\beta (> 2.5)$. The UCD/dE,N includes 
two compact sources with hints of extended envelopes.  We also plot 3 individual bright UCDs (black stars) but do
not include the large errorbars.
\label{metals}}       
\end{figure}

Stellar population abundances and ages are measured using the \citet{schiavon} models.  We find that 
individual UCD spectra have too low S/N to measure abundances accurately.  We therefore combine 
spectra of objects with similar 
color, distribution, or spectral properties (Figure ~\ref{metals}).   Although errorbars remain large 
even for the composite spectra, we find that UCDs display a large range in ages and metallicities, significant
at $> 2.5\sigma$.  Blue UCDs ($V - I < 1.05$) are found to have intermediate ages and metal 
poor populations, while red UCDs may have older ages and have much higher metallicities.  
There is some evidence for UCDs with very young populations.  However, the UCD shown in the plot with
a young age appears to have a faint extended envelope and may not even be in the same class of object
as the other UCDs.  The one cE for which
we have two abundance measurements is both more metal rich and younger than the UCDs.
Other cEs, from \citet{compact}, have metallicities as low as those found for the red
UCDs.

\subsection{Distribution}
\label{sec:1b}

The UCDs have thus far been found exclusively in the central core region, along a fairly
narrow band in declination.  This apparent structure through the core region may be due to
small number statistics, although considering the full sample of good candidates (green points
in Figure ~\ref{fig1}), we find very few candidates lying far north or south of this band.  A strong
concentration of confirmed and candidate UCDs is also seen around the cD galaxy NGC 4874.
Fewer UCDs are found around NGC 4889, the other central giant, but this is possibly due to a gap in our
ACS coverage at this location.

\begin{figure}
\resizebox{1.05\columnwidth}{!}{\includegraphics{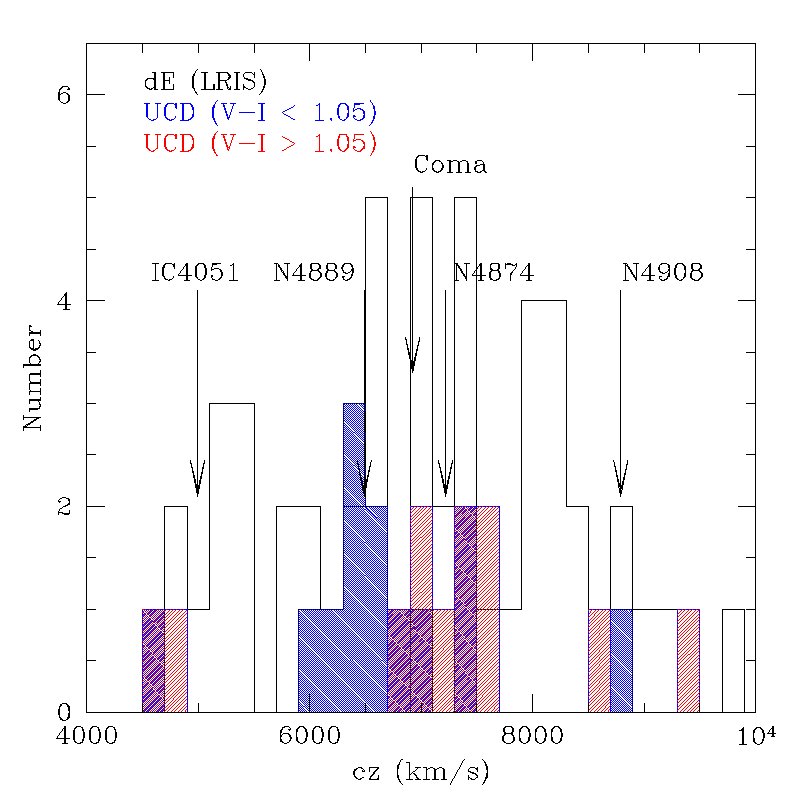}}
\caption{Histograms of radial velocities for UCDs separated by $V-I$ color.
\label{colhist}}       
\end{figure}

The mean radial velocity for the full set of confirmed UCDs is centered on the
Coma cluster mean.  A velocity dispersion of $\sigma = 1072$ km/s for the UCDs is insignificantly
lower than what we find for dEs in the same region.   Velocity
histograms are shown in Figure ~\ref{colhist}, with separation by UCD $V-I$ color.  Peculiar velocities
of prominent cluster giants  are indicated.  A large fraction of UCDs
have velocities close to that of the cluster mean although there are a few outliers
which have velocities similar to other cluster giants.  When the UCDs are separated by
color, they appear to split into two groups centered around the two central dominant
giants.  If we consider only UCDs with RA $< 195$, we find 
$\langle v_r\rangle = 7296\pm160$ km/s with $\sigma = 558\pm114$ km/s, while 
UCDs east of this have $\langle v_r\rangle = 6559\pm325$ km/s and $\sigma = 1257\pm230$ km/s.
The differences are significant at the $\sim 3\sigma$ level.  

\begin{figure}
\resizebox{1.05\columnwidth}{!}{\includegraphics{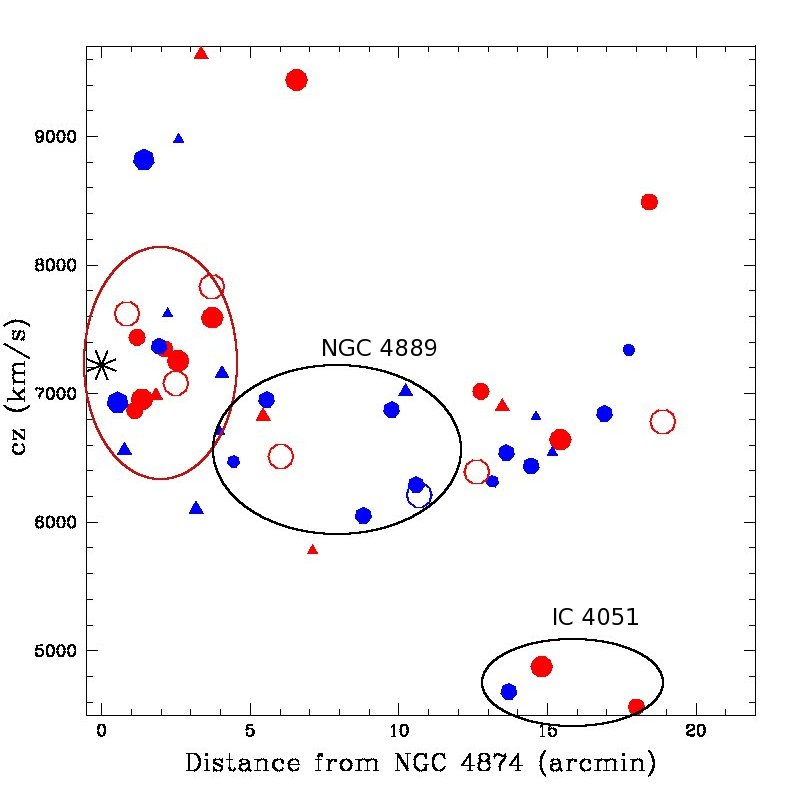}}
\caption{UCD radial velocity as a function of projected distance from 
cD galaxy NGC 4874 (asterisk).   UCDs with radial velocities similar to
this galaxy with small separations in projected distance are encircled in red.
UCDs possibly associated
with the cluster giants NGC 4889 and IC 4051 (a galaxy with one
of the highest Coma Cluster globular cluster specific frequencies) are also
encircled. 
One arcmin corresponds to 29 kpc.
Red symbols have $(V-I) > 1.05$, while blue have $(V-I) < 1.05$. Where $V-I$ colors do
not exist, we use colors in other bands to infer a rough $V-I$ color. Larger
symbols represent brighter magnitudes.  Open circles are cEs, triangles are
objects with unsecure redshifts. 
\label{assoc}}       
\end{figure}

\begin{figure}
\resizebox{1.05\columnwidth}{!}{\includegraphics{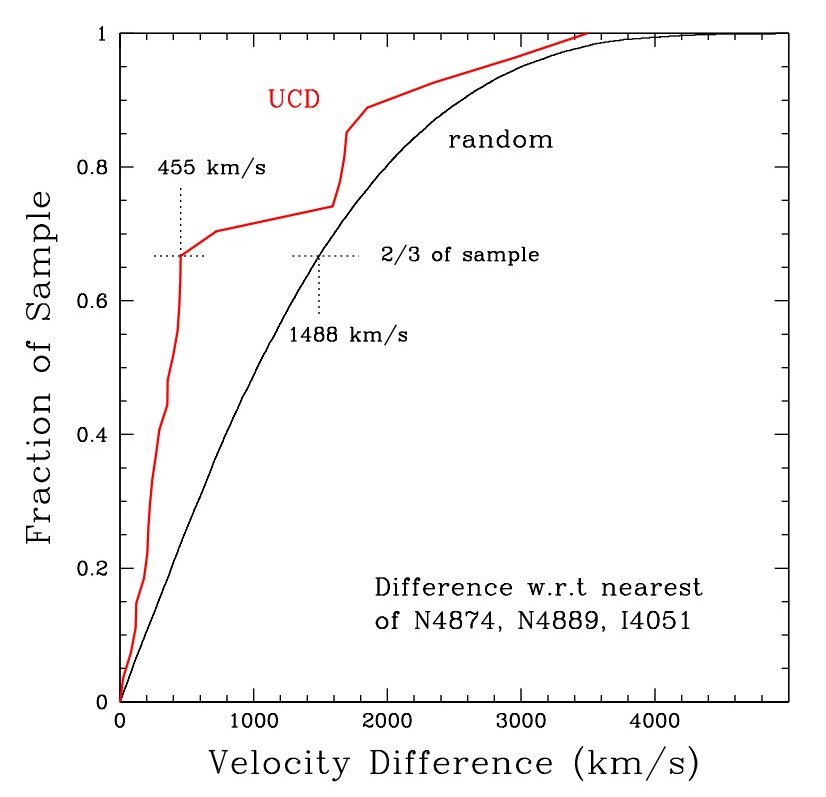}}
\caption{Cumulative distribution of
the velocity difference between each UCD and the nearest of one of 3 giants in
the core region - NGC 4874, NGC 4889, and IC 4051 (red
line). The black curve displays the same for 25,000 points randomly
distributed over the LRIS footprint and attached to randomly sampled
velocities assuming a Gaussian distribution with
 $\langle v_r\rangle = 6925$ km/s and $\sigma = 1000$ km/s.
\label{cumdist}}       
\end{figure}

Of greater import, when we take into account spatial proximity with the 
central giants, we find 9 UCDs near NGC 4874 with 
$\langle v_r\rangle = 7257\pm87$ km/s and $\sigma = 254\pm60$ km/s and
5 UCDs near NGC 4889 with  
$\langle v_r\rangle = 6526\pm148$ km/s and $\sigma = 332\pm105$ km/s.  The
central giants NGC 4874 and NGC 4889 have peculiar velocities of
7220 and 6495 km/s 
respectively.  We display possible associations of UCDs with three prominent giants in Figure ~\ref{assoc}.
In Figure ~\ref{cumdist}, we show the cumulative distribution of confirmed
UCDs as a function of velocity difference with respect to the nearest of
one of the 3 giants: NGC 4874, NGC 4889, and IC 4051.  This is compared to
the expectation for a spatially uniform distribution of objects having a Gaussian 
velocity distribution assuming the mean Coma cluster parameters.  We find
that 2/3 of the UCD sample exhibit a much stronger concentration around 
cluster giants than would be expected for a random distribution.
Furthermore, based on the escape velocities and tidal radii for the 2 central giants,
we expect that the UCDs which are coincident in space and velocity are bound
to these galaxies.  Similar isolation in space and velocity around other
giants are noted, although in these cases, it is more difficult to disentangle the UCDs
from the cluster potential.

\section{Discussion} \label{sec:2}
Overall we find the properties and distribution of the Coma cluster UCDs to be very similar to what has
already been found for UCDs in other environments.  Sizes, magnitudes, colors, metallicities, and even 
a distribution exhibiting strong associations with individual galaxies all
agree with previous findings for UCDs.

\begin{figure}
\resizebox{1.05\columnwidth}{!}{\includegraphics{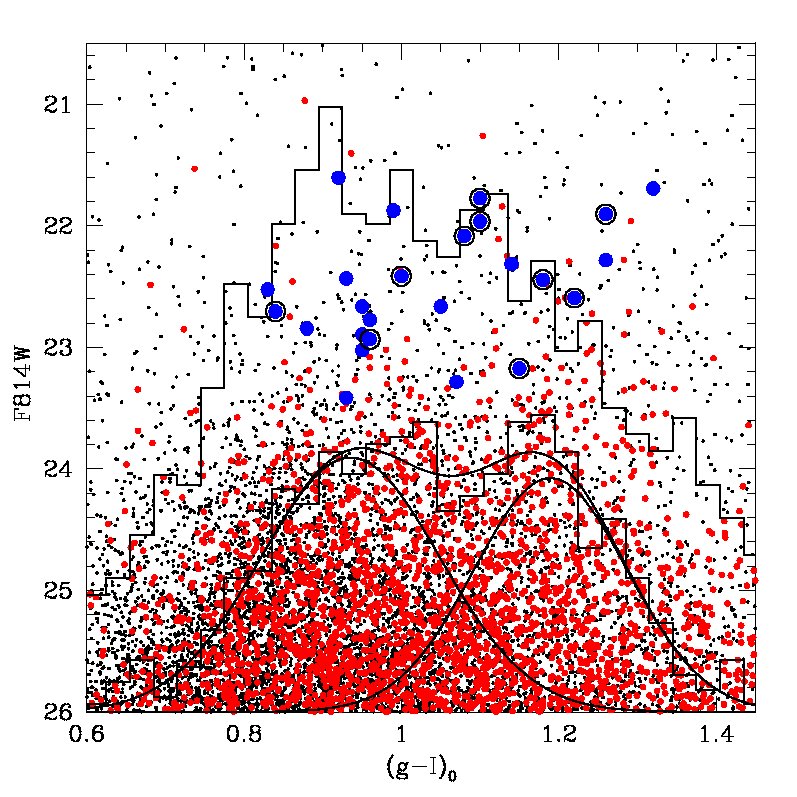}}
\caption{Color distribution of Coma cluster candidate globular clusters
(black points: ACS survey region, red points: visit 19 which includes NGC 4874) and 
confirmed UCDs (all: blue, circled: visit 19 only).
The upper histogram displays binned $F814W > 24.7$ visit 19 globular cluster
counts, the lower histogram includes only $F814W < 24.7$ visit 19 counts.  We show
the best double gaussian fit to this set of brighter counts. The
two peaks are at $g - I = 0.935$ and 1.189.  Data for the globular clusters 
comes from \citet{peng09}.
\label{gccolor}}       
\end{figure}

A comparison with the Coma Cluster globular cluster population hints at one potential origin.
In Figure ~\ref{gccolor} we compare the color-magnitude diagram for GCs with our confirmed UCDs.
We note a similar and very broad color spread for the UCDs, something not observed for normal dwarf galaxies.
We also find no evidence for a discontinuity over the luminosity range between GCs and UCDs.  In
fact, due to this lack of any distinction in luminosity space, most of our UCDs are included in the
GC list of \citet{peng09}.  The continuous extension of GC luminosities into the range of UCDs, in particular
for the red branch of GCs, had been previously noted for compact stellar systems around the
cD galaxy NGC 3311 \citep{wehner08}.  

Comparing to the spatial distribution of GCs from \citet{peng09}, we
find a concentration of UCDs around NGC 4874 which also has a very high GC specific frequency.
In addition we see a hint of an
excess of UCD candidates towards IC 4051, which boasts one of the highest GC specific
frequencies in the Coma cluster. Unfortunately, this galaxy lies just east of our ACS footprint.  
Confirmation of a large population of UCDs around this galaxy
would provide strong support for a star cluster origin.
\citet{peng09} also note a potential excess of GCs running E - W
in a band through the core, in the same region where presently all confirmed UCDs reside,
and where the majority of our best candidates are found (Figure ~\ref{fig1}).

Given this strong correlation with globular cluster properties and similar spatial distributions, we can
argue that UCDs are consistent with simply constituting a bright extension of the globular cluster sequence.
Arguments against this scenario have included the larger sizes found for UCDs and the discovery in a few
UCDs of higher M/L ratios than can be
explained with canonical IMFs and baryonic matter \citep{has05}.  This would appear to invoke a
need for dark matter, more in line with a galaxy rather than pure star cluster origin. 
However, recent work by \citet{murray09} and \citet{dab09} find an explanation in terms of
the physics at the time of formation of the star cluster which can account both for the larger sizes
of UCDs and for the higher M/L ratios assuming a top heavy IMF.

The red UCDs are the most strongly clustered population and are primarily associated with the
cD galaxy NGC 4874.  This is perhaps expected since more massive galaxies tend to host both more massive and
more metal rich globular clusters \citep{peng06,harris09,hilker09}.  Some of the blue UCDs can be identified with
specific giant galaxy hosts, while others may lie within the intracluster region.  These
lie along a band exactly where an excess of GCs has been noted \citep{peng09}.   It is
possible these are threshed nuclei of the same dwarfs who lost their globular clusters to the
Coma cluster potential, although it is likewise possible that they are themselves simply stripped star clusters
from giant galaxies.

In summary, although we cannot rule out the galaxy threshing hypothesis, we do find some support for 
a GC-UCD relationship.
Besides having properties and a distribution similar to that of the Coma GCs,
we find strong spatial and velocity correlations with the
major galaxies and show that a majority of the confirmed UCDs reside within the halos of those
galaxies.  The UCDs furthermore exhibit color and metallicity correlations with at least
some of the giant galaxies. NGC 4874 hosts a predominantly red population, while NGC 4889 is currently
associated with only blue UCDs, although a radial gradient cannot be ruled out due to a missing
observation centered on this galaxy.  This differentiation in color by host suggests
formation in discrete star formation events, from pre-enriched host environments.  It is difficult
to imagine how threshing of dE,N could produce a similar color/metallicity distribution.

A couple of the confirmed UCDs do show evidence for having extremely low surface brightness
envelopes and may be entirely different entities.  We also note that we find some similarities
with the cE population, but a 1.5 magnitude gap currently separates these two populations.
The majority of the UCDs are consistent with a star cluster origin.


%
\bibliographystyle{epj}


\end{document}